\pdfoutput=1
\documentclass[aps, prd
, twocolumn
, nofootinbib 
, longbibliography
, floatfix
]{revtex4-1}
\usepackage{graphicx}
\usepackage[caption=false]{subfig}
\usepackage{mathrsfs}
\usepackage{amsmath,amssymb,amsthm}
\usepackage{bm}
\usepackage{braket}
\usepackage{listings}
\usepackage{cases}
\usepackage{comment}
\usepackage{soul}
\usepackage{cancel}
\usepackage{cases}
\usepackage[utf8]{inputenc}
\usepackage{url}
\usepackage{longtable}
\usepackage[normalem]{ulem}
\usepackage[colorlinks=true
,urlcolor=blue
,anchorcolor=blue
,citecolor=blue
,filecolor=blue
,linkcolor=blue
,menucolor=blue
,linktocpage=true
,pdfproducer=medialab
,pdfa=true
]{hyperref}

\newcommand{\dif}[2]{\frac{\mathrm{d} #1}{\mathrm{d} #2}}

\newcommand{\dd}{\mathrm{d}}

\newcommand{\ee}{\mathrm{e}}

\newcommand{\Mpl}{M_\text{Pl}}
\newcommand{\ns}{n_{{}_\mathrm{S}}}

\renewcommand{\Re}{\mathrm{Re}}

\newcommand{\eff}{\text{eff}}

\newcommand{\thr}{\text{th}}
\newcommand{\eq}{\text{eq}}

\newcommand{\eg}{\textit{e.g. }}
\newcommand{\etal}{\textit{et al. }}
\newcommand{\ie}{\textit{i.e. }}

\newcommand{\calC}{\mathcal{C}}

\newcommand{\calH}{\mathcal{H}}

\newcommand{\calO}{\mathcal{O}}

\newcommand{\calP}{\mathcal{P}}
\newcommand{\calR}{\mathcal{R}}

\newcommand{\calT}{\mathcal{T}}

\newcommand{\bae}[1]{\begin{align} #1 \end{align}}
\newcommand{\bce}[1]{\begin{cases} #1 \end{cases}}
\newcommand{\dps}{\displaystyle}
\newcommand{\bfe}[4]{
\begin{figure} 
	\centering
	\includegraphics[#1]{#2}
	\caption{#3}
	\label{#4}
\end{figure}}

\newtheorem*{swampland}{Redefined de Sitter Conjecture}

\def\mnras{Mon. Not. Roy. Astron. Soc.}             % Monthly Notices of the RAS

\begin{document}
\title{Primordial black hole tower: Dark matter, earth-mass, and LIGO black holes}
\date{\today}

\author{Yuichiro Tada}
\email{tada.yuichiro@e.mbox.nagoya-u.ac.jp}
\affiliation{Department of Physics, Nagoya University, Nagoya 464-8602, Japan}
\author{Shuichiro Yokoyama}
\email{shu@kmi.nagoya-u.ac.jp}
\affiliation{Kobayashi Maskawa Institute, Nagoya University, Aichi 464-8602, Japan}
\affiliation{Kavli IPMU (WPI), UTIAS, The University of Tokyo, Kashiwa, Chiba 277-8583, Japan}

\begin{abstract}
We investigate a possibility of primordial black hole (PBH) formation with a hierarchical mass spectrum in
multiple phases of inflation. As an example, 
we find that one can simultaneously realize 
a mass spectrum which has recently attracted a lot of attention:
stellar-mass PBHs ($\sim\mathcal{O}(10)M_\odot$) as a possible source of binary black holes detected by LIGO/Virgo collaboration, asteroid-mass ($\sim\mathcal{O}(10^{-12})M_\odot$) as
a main component of dark matter, and earth-mass ($\sim\mathcal{O}(10^{-5})M_\odot$) as a source of ultrashort-timescale events in Optical Gravitational Lensing Experiment microlensing data.
The recent refined de Sitter swampland conjecture may support such a multi-phase inflationary scenario with hierarchical mass PBHs as a transition signal of each inflationary phase.
\end{abstract}

\maketitle
%\tableofcontents

\section{Introduction}

The epochal success of first direct detection of gravitational waves (GWs) GW150914 by the LIGO/Virgo collaboration~\cite{Abbott:2016blz} opens up a quite new research field of the universe.
Their subsequent observations so far reveal the ubiquitous $\sim\calO(10)M_\odot$ 
black holes (BHs) which are more massive than one thought.
For the origin of such massive stellar BHs, one might have recourse to primordial ones. Several cosmological scenarios predict BH formation in advance of ordinary stars, which is called primordial black holes (PBHs). 
For example, an order-unity overdense region can collapse to a BH by self-gravitation soon after its horizon reentry during the radiation-dominated era~\cite{1971MNRAS.152...75H,Carr:1974nx,Carr:1975qj}. 

Right after GW150914, the possibility of PBHs being GW sources were discussed~\cite{Bird:2016dcv,Clesse:2016vqa}\footnote{It has been also reported that the Poissonian isocurvature density perturbations associated with $\calO(10)M_\odot$ PBHs might be a possible source of the observed cosmic infrared background fluctuations~\cite{Kashlinsky:2016sdv}.} and it was suggested that even PBHs comprising only sub-percent of total dark matters (DMs) can form binaries during the radiation-dominated era and explain the estimated current merger rate~\cite{Sasaki:2016jop}.
Making up all DMs by PBHs itself is also an interesting scenario other than GW.
Though such a scenario had been thought to be ruled out observationally, several authors recently revisited the constraints by gravitational lensing events, one of the main constraining schemes on PBH abundance particularly for light mass~\cite{Inomata:2017vxo,Katz:2018zrn}.
They showed that the finite size of luminous sources or the wave property of photons significantly reduces the lensing efficiency
and now windows for all DMs are open on around asteroid-mass $\sim10^{-16}\text{--}10^{-14}M_\odot$ and $\sim10^{-13}\text{--}10^{-11}M_\odot$.
Furthermore, Ref.~\cite{Niikura:2019kqi} has recently shown that 
earth-mass PBHs $\sim10^{-5}M_\odot$ comprising $\calO(1)\%$ of total DMs can explain
ultrashort-timescale microlensing events which are reported in
5-year observations of stars in Galactic bulge by the Optical Gravitational Lensing Experiment (OGLE)~\cite{2017Natur.548..183M}.\footnote{This mass range might be also interesting in the context of the superradiant phenomena
of the axion~\cite{Rosa:2017ury}. If PBHs are spinning up by their mergers, the superradiant instability on the QCD axion will form its cloud surrounding PBHs and then the stimulated axion decay gives rise to extremely bright
lasers which might explain the observed fast radio bursts.}
Thus, it might be interesting to investigate the possibility of generating such a hierarchical PBH mass spectrum in the early universe.

One of the widely studied sources of the overdense region collapsing to a BH in the early universe is
large primordial curvature perturbations, which could be realized by inflation like as those on the cosmic microwave background (CMB) scale.
However the CMB observations indicate their amplitudes $\sim10^{-5}$~\cite{Aghanim:2018eyx} which are far from order-unity, and moreover
standard single-field slow-roll inflation models predict almost scale-invariant perturbations in general. 
Thus in order to realize large primordial perturbations to form PBHs,
one needs much amplification on smaller scales by considering extended inflationary models (see \eg a recent review article~\cite{Sasaki:2018dmp}).
Among such inflationary models, double-phase inflation has attracted attention recently~\cite{Kawasaki:2016pql,Inomata:2016rbd,Inomata:2017okj,Inomata:2017uaw,Inomata:2017vxo}, which can yield PBH formation with nearly monochromatic mass spectrum as a transition signal between the two inflationary phases.\footnote{Similarly, the PBH formation in continuous-phase double inflation, \textit{i.e.}, hybrid inflation has been also studied well (see \eg \cite{GarciaBellido:1996qt,Clesse:2015wea,Kawasaki:2015ppx}). The resultant PBH mass spectrum in this case tends to be too broad to avoid the observational constraints, compared to our double/multi-phase ones.}
From this viewpoint, the hierarchy in preferred PBH masses would suggest multi-phase (more than double) inflation.
This possibility is important also in the context of string theory.
According to the refined de Sitter swampland conjecture~\cite{Obied:2018sgi,Garg:2018reu,Ooguri:2018wrx}, even quasi-stable de Sitter vacua might not be consistent with 
UV-complete effective field theories, implying each continuous phase of inflation cannot last so long, compared to required e-folds $\sim50\text{--}60$ for our whole observable universe. A series of short inflationary phases instead can account for total e-folds, leaving some signals of phase transitions.
In fact our model satisfies the second condition of the refined de Sitter conjecture except in the CMB phase, for which we do not assume any specific model in this paper.

We investigate in this paper a PBH formation scenario with a hierarchical mass spectrum in multi-phase inflation. In Sec.~\ref{sec: PBH}, we review the estimation procedure of the PBH abundance as well as the current observational constraints. The resultant PBH mass spectrum in our model is also shown in Fig.~\ref{fig: fPBH} first.
Our multi-phase inflationary model is described in detail in Sec.~\ref{sec: inflation}.
As its testability, our model leaves stochastic GWs by the second-order effect of large primordial scalar perturbations with sufficient amplitude to be detected by future observations as discussed in Sec.~\ref{sec: GW}.
Sec.~\ref{sec: conclusions} is devoted to the conclusion as well as the discussion about the swampland conjecture.
The procedure dependence of the PBH abundance is mentioned in Appendix~\ref{sec: uncertainties}.
We adopt the natural unit $c=\hbar=1$ throughout this paper.

\section{Hierarchical primordial black hole spectrum}\label{sec: PBH}

\subsection{Primordial black holes for dark matters, LIGO/Virgo GW, and OGLE ultrashort-timescale microlensing events}

An extreme astrophysical object, black hole (BH), has been playing a key role 
not only in astrophysics, but also in gravitational theory, quantum physics, 
particle physics and cosmology.
While it forms as a remnant of an explosive death of a massive star in general,
several cosmological scenarios can also predict abundant BH formation in the early universe before the ordinary star formation, called \emph{primordial black hole} (PBH).
For a PBH formed in the radiation-dominated era, its mass is roughly given by the horizon mass at its formation time:
\bae{
    M_H\sim\frac{t}{G}\sim M_\odot\left(\frac{t}{10^{-5}\,\mathrm{s}}\right),
}
where $G\simeq6.7\times10^{-39}\,\mathrm{GeV}^{-2}$ is the Newtonian constant of gravitation and
$M_\odot\simeq2\times10^{33}\,\mathrm{g}$ denotes the solar mass.
It shows that the possible mass of PBH spans a very wide range, from the Planck mass $\simeq2\times10^{-5}\,\mathrm{g}$ as the extremely light end to \eg $\sim10^5M_\odot$ for seeds of supermassive BHs at galactic cores~\cite{Bean:2002kx,Carr:2018rid}. Particularly if a BH lighter than the sun is found, it should be a primordial one instead of the ordinary remnant of star.

While ones lighter than $\sim10^{15}\,\mathrm{g}$ have evaporated by now, more massive PBHs can survive against the Hawking radiation and play a role of the main/sub component of dark matters (DMs).
Their abundance has been constrained by many kinds of observations represented by gravitational lensing events, depending on their mass. The current conservative constraints are summarized in Fig.~\ref{fig: fPBH}. 
It should be noted that PBHs of $\sim10^{-16}\text{--}10^{-11}M_\odot$ had been thought to be constrained by non-detection of femtolensing events of gamma-ray bursts~\cite{2012PhRvD..86d3001B} or microlensing against the Andromeda galaxy~\cite{Niikura:2017zjd}.
However it was recently reported that the finiteness of the luminous source size and the wave property of photons significantly reduce the lensing efficiency~\cite{Niikura:2017zjd,Inomata:2017vxo,Katz:2018zrn}, and the windows for $\sim10^{-16}\text{--}10^{-14}M_\odot$ and $\sim10^{-13}\text{--}10^{-11}M_\odot$ are open now. Therefore PBHs can comprise all DMs in this range.

\bfe{width=\hsize}{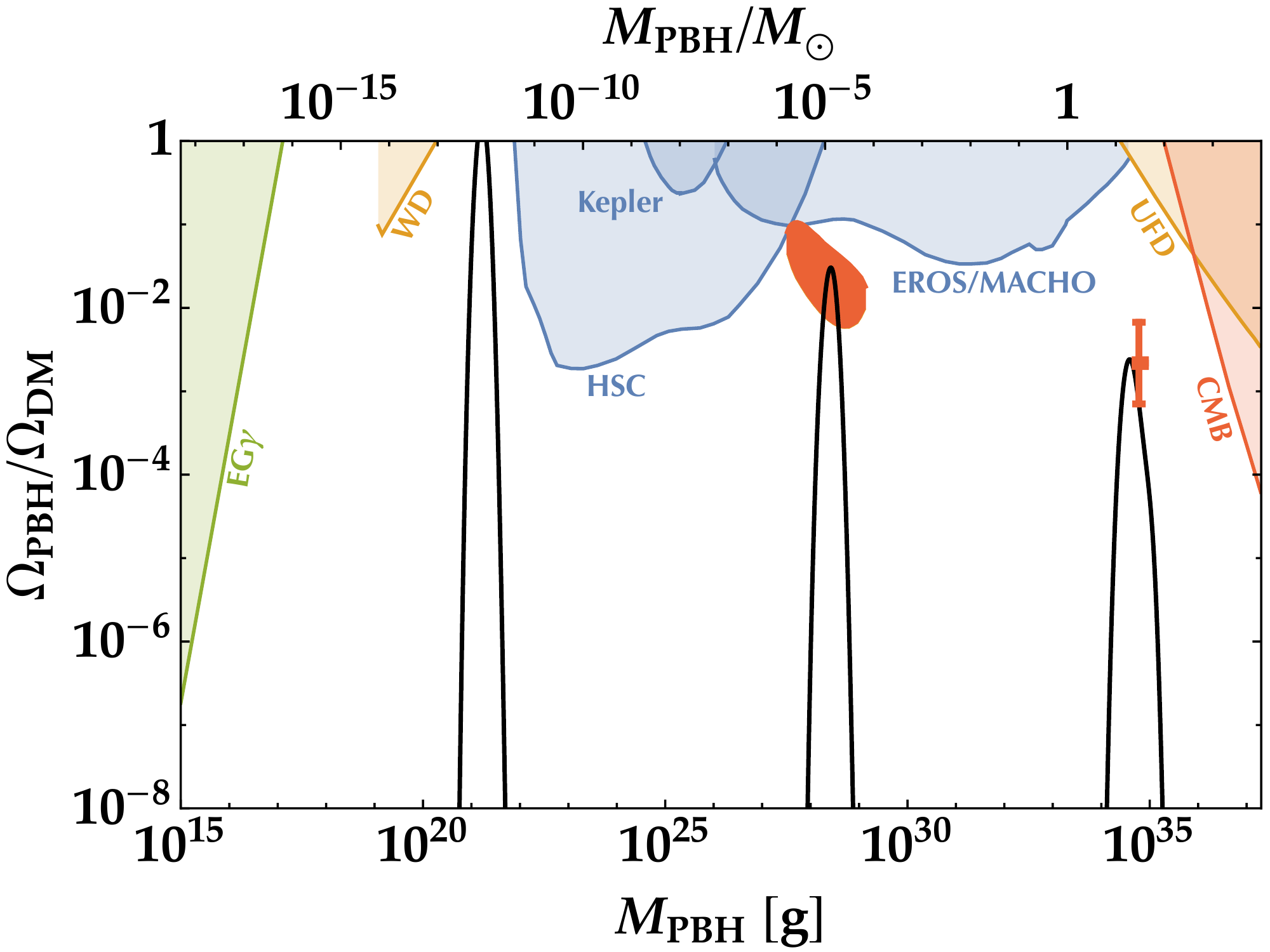}{The predicted PBH mass spectrum is shown by the black lines. The lines with half shades represent the current observational constraints: extra-galactic gamma-ray by the Hawking radiation (EG$\gamma$)~\cite{Carr:2009jm}, non-destruction of white dwarfs in our local galaxy (WD)~\cite{Graham:2015apa}, Subaru HSC microlensing (HSC)~\cite{Niikura:2017zjd}, Kepler milli/microlensing (Kepler)~\cite{Griest:2013esa},
EROS/MACHO microlensing (EROS/MACHO)~\cite{Tisserand:2006zx}, dynamical heating of ultra-faint dwarf galaxies (UFD)~\cite{Brandt:2016aco}, and the most conservative accretion constraints by CMB (CMB)~\cite{Ali-Haimoud:2016mbv,Blum:2016cjs,Horowitz:2016lib,Poulin:2017bwe}. The red region is the inferred PBH abundance by the OGLE ultrashort-timescale microlensing events~\cite{Niikura:2019kqi}, 
while the point with error bars corresponds with GW150914~\cite{Sasaki:2016jop}.} {fig: fPBH}

As other interesting regions, it has been suggested that PBHs can form their binaries by the gravitational attracting force between them even during the radiation-dominated era and the merger rate estimated by LIGO/Virgo gravitational events can be explained if the PBH fraction to total DM is $\sim0.1\%$~\cite{Sasaki:2016jop} as indicated by a red point with error bars for GW150914 event~\cite{Abbott:2016nhf} in Fig.~\ref{fig: fPBH}.\footnote{We focus only on the first event GW150914 with the approximation of the monochromatic mass function as a simplified indicator for brevity though in total ten binary-BH events with slightly varied masses have been reported so far~\cite{LIGOScientific:2018mvr}. The error in the PBH abundance corresponds with the range of the merger rate $2\text{--}53\,\mathrm{Gpc}^{-3}\mathrm{yr}^{-1}$, while that in the PBH mass is given by the high end of the massive one $(36+5)M_\odot$ and the low end of the lighter one $(29-4)M_\odot$ respectively~\cite{Abbott:2016nhf}.
For a refined estimation with an extended mass function including all ten events, see \eg Refs.~\cite{Raidal:2017mfl,Raidal:2018bbj}.}
LIGO/Virgo collaboration reported that they detected ten GW events from binary BH mergers during the first and second observing runs~\cite{LIGOScientific:2018mvr}, and
masses of observed BHs $\sim30\text{--}50M_\odot$ are slightly more massive than previously found stellar ones. Thus, it would suggest that PBHs can be a candidate of these events 
instead of the astrophysical ones.
In addition to this range, 
recently, 
Ref.~\cite{Niikura:2019kqi} investigated the possibility that
earth-mass PBHs can explain ultrashort-timescale microlensing events
which has been reported in 5-year observations of million stars in the Galactic bulge fields
by the Optical Gravitational Lensing Experiment (OGLE) collaboration,
without introducing free-floating planets~\cite{2017Natur.548..183M}.
According to Ref.~\cite{Niikura:2019kqi}, the required abundance is $\sim1\%$ to total DMs as indicated by the red region in Fig.~\ref{fig: fPBH}.

\subsection{Mass spectrum of primordial black holes}

In order to make a theoretical estimation for the required PBH mass spectrum 
as discussed above, one needs to specify the formation scenario of PBH.
Amongst several scenarios such as 
the collapse of topological defects~(see \eg Ref.~\cite{Deng:2016vzb})
or the gravitational growth of density perturbations enhanced by disappearance of fluid pressure~(see \eg Ref.~\cite{Jedamzik:1996mr}),
we focus on a widely-studied one that order-unity overdense Hubble patches crunch into BHs during the radiation-dominated era.
As discussed in the pioneer work by Carr and Hawking~\cite{Carr:1974nx}, a sufficiently overdense Hubble patch is expected to be decoupled from the background expansion beyond the Jeans scale and collapse directly into a BH soon after its horizon reentry.
The threshold for collapse was first briefly estimated by Carr~\cite{Carr:1975qj} as $\delta_\thr\sim w$ at the horizon reentry on the uniform Hubble slice where $w=p/\rho=1/3$ is the equation of state for radiation fluid, verified by several numerical simulations as $\delta_\thr\simeq0.4$~\cite{Shibata:1999zs,Musco:2004ak} on the comoving slice, and also theoretically refined by Harada, Yoo and Kohri~\cite{Harada:2013epa} as $\delta_\thr=\frac{3(1+w)}{5+3w}\sin^2\frac{\pi\sqrt{w}}{1+3w}\simeq0.4$.
Though these values slightly depend on the profile of density perturbations~(see \eg Ref.~\cite{Nakama:2013ica}),
we adopt $\delta_\thr=0.4$ as a fiducial value in this paper.

Based on the Press-Schechter approach which is a conventional way to evaluate the mass spectrum of collapsed objects,
with an assumption that the primordial density perturbations $\delta$ on comoving slice follow the Gaussian distribution for simplicity (see \eg Refs.~\cite{Kawasaki:2019mbl,DeLuca:2019qsy,Young:2019yug} for the non-Gaussian effect due to the non-linear relation between the density and curvature perturbations), 
the formation probability $\beta(R)$ of collapsed objects (\ie PBHs) on some coarse-graining comoving scale $R$ is given by
\bae{\label{eq: betaR}
    \beta(R)&=\int_{\delta_\thr}\frac{1}{\sqrt{2\pi\sigma_\delta^2(R)}}\exp\left[-\frac{\delta^2}{2\sigma_\delta^2(R)}\right]\dd\delta \nonumber \\
    &=\frac{1}{2}\mathrm{erfc}\left(\frac{\delta_\thr}{\sqrt{2\sigma_\delta^2(R)}}\right).
}
Here the variance $\sigma_\delta^2(R)$ is related with the initial power spectrum  of the conserved curvature perturbations on comoving slice, $\calP_\calR(k)$, as 
\bae{\label{eq: sigma_delta}
    \sigma_\delta^2(R)\!=\!\int\!\dd\log k\,\tilde{W}^2(kR)\frac{16}{81}(kR)^4T^2(k,\eta=R^{-1})\calP_\calR(k),
}
where $T(k,\eta)$ represents the scalar transfer function during the radiation-dominated era given by
\bae{
    T(k,\eta)=3\frac{\sin(k\eta/\sqrt{3})-(k\eta/\sqrt{3})\cos(k\eta/\sqrt{3})}{(k\eta/\sqrt{3})^3}.
}
For the window function $\tilde{W}$, we adopt the real-space top-hat one in Fourier space:
\bae{
	\tilde{W}(kR)=3\frac{\sin kR-kR\cos kR}{(kR)^3}.
}

The mass of formed PBH is related with the coarse-graining scale $R$ by the horizon mass at its horizon reentry $aH=R^{-1}$ as
\bae{\label{eq: MR}
    M(R)&=\left.\gamma\rho\frac{4\pi}{3}H^{-3}\right|_{aH=R^{-1}} \nonumber \\
    &\simeq\frac{\gamma M_\eq}{\sqrt{2}}\left(\frac{g_{*\eq}}{g_*(T_R)}\right)^{1/6}\left(Rk_\eq\right)^2
    \nonumber \\
    &\simeq10^{20}\gamma\left(\frac{g_*(T_R)}{106.75}\right)^{-1/6}\left(\frac{R}{6.4\times10^{-14}\,\mathrm{Mpc}}\right)^2\,\mathrm{g},
}
where $M_\eq=\frac{8\pi}{3}\frac{\rho^r_0}{a_\eq k_\eq^3}$ is the horizon mass at the matter-radiation equality with the current radiation density $\rho^r_0\simeq7.84\times10^{-34}\,\mathrm{g}\,\mathrm{cm}^{-3}$ as well as the comoving horizon scale $k_\eq\simeq0.07\Omega_mh^2\,\mathrm{Mpc}^{-1}$ and the scale factor $a_\eq\simeq(2.4\times10^4\Omega_mh^2)^{-1}$ at the equality~\cite{Josan:2009qn}.
$\Omega_mh^2\simeq0.143$~\cite{Aghanim:2018eyx} is the current density parameter for the matter components.
$g_*$ denotes the effective degrees of freedom for energy density ($g_{*\eq}=3.38$ at the equality), assumed to be almost equal to those for entropy density. 
It is a function of the fluid temperature (see \eg Ref.~\cite{Aghanim:2018eyx}), while the coarse-graining scale $R$ and the temperature $T_R$ can be related by~\cite{Inomata:2018epa}
\bae{
    (Rk_\eq)^{-1}\simeq2(\sqrt{2}-1)\left(\frac{g_*}{g_{*\eq}}\right)^{1/6}\frac{T_R}{T_\eq},
}
with the temperature $T_\eq=\frac{2.725\,\mathrm{K}}{a_\eq}$ at the equality.
$\gamma$ is a numerical parameter representing the mass efficiency of collapse.
In this paper $\gamma=1$ is adopted for simplicity though there are several works addressing this value (\eg Ref.~\cite{Musco:2004ak}).

After their formation, PBHs behave as non-relativistic matters and therefore their current fraction to total cold DMs is given by
\bae{\label{eq: fPBH}
	&f_\mathrm{PBH}(M)=\frac{\Omega_\mathrm{PBH}(M)h^2}{\Omega_\mathrm{DM}h^2}=\frac{T_R}{T_\eq}\frac{\Omega_mh^2}{\Omega_\mathrm{DM}h^2}\gamma\beta(R) 
	\nonumber \\
	&\simeq\!\gamma^\frac{3}{2}\!\left(\!\frac{\beta(R)}{7.2\times10^{-16}}\!\right)\!\left(\!\frac{\Omega_\mathrm{DM}h^2}{0.12}\!\right)^{\!-1}\!\left(\!\frac{g_*(T_R)}{106.75}\!\right)^{\!-\frac{1}{4}}
	\!\left(\!\frac{M}{10^{20}\,\mathrm{g}}\!\right)^{\!-\frac{1}{2}}\hspace{-5pt}.
}
$\Omega_\mathrm{DM}h^2\simeq0.12$~\cite{Aghanim:2018eyx} is the current density parameter of total DMs.
As mentioned, the estimation procedure of the PBH abundance involves several uncertainties in the model parameters, the choice of the window function, and so on. Given the PBH abundance fixed, such uncertainties are inherited by the required primordial perturbations and affect model testability described in Sec.~\ref{sec: GW} as discussed in Ref.~\cite{Ando:2018qdb}.
The refined procedure in the peak theory has been also proposed recently beyond the Press-Schechter approach~\cite{Yoo:2018esr,Germani:2018jgr}. 
We address these issues in Appendix~\ref{sec: uncertainties}.

One finds from these equations that sizable amount of PBHs requires the significant amplification of the primordial curvature perturbations as $\calP_\calR\simeq10^{-2}$
on the corresponding scale (\eg $k\sim R^{-1} \sim 10^{12}~{\rm Mpc}^{-1}$ for DM-PBH with $\sim10^{20}\,\mathrm{g}$) compared to $\calP_\calR\simeq2\times10^{-9}$ on the cosmic microwave background (CMB) scale $k_*\sim0.05\,\mathrm{Mpc}^{-1}$~\cite{Aghanim:2018eyx}.
However both the scale dependence of $\calP_\calR$, \ie $\ns-1=\dif{\log\calP_\calR}{\log k}$, and its running $\alpha=\dif{\ns}{\log k}$ are slow-roll suppressed in the simplest
single-field slow-roll inflation, prohibiting such an amplification. This difficulty has been reported even beyond the slow-roll approximation~\cite{Motohashi:2017kbs,Passaglia:2018ixg}.
On the other hand, if one allows two/multiple phases of inflation during the last $50\text{--}60$ e-folds,
the PBH scale can be free from the CMB scale constraint and such an amplification can be realized much more easily.
In particular, Refs.~\cite{Kawasaki:2016pql,Inomata:2016rbd,Inomata:2017okj,Inomata:2017uaw,Inomata:2017vxo} show that sharp peaks of the curvature power spectrum can
appear on the scales corresponding with the transitions between inflationary phases.
From this viewpoint, each interesting mass region mentioned above may correspond with such an transition time of some multi-phase inflation.
In the next section, a concrete multi-phase model is shown which realizes the three peaks of the PBH mass spectrum simultaneously as shown in Fig.~\ref{fig: fPBH}.

\section{Multi-phase inflation}\label{sec: inflation}

In this section we consider a quadruple-phase inflationary model to realize three peaks in the PBH mass spectrum at each phase transition.
Note that we do not specify the model responsible for the CMB scale
or the reheating mechanism but simply assume that they consistently succeed.
Each inflationary phase is governed by a different scalar inflaton $\phi_i$, $(i=1,2,3,4)$.
As the energy scales for later phases should be lower than those for earlier phases and that on the CMB scale is already constrained from above as 
$\rho_\mathrm{inf,CMB}\lesssim (1.7 \times 10^{16}~\mathrm{GeV})^4$ by the non-detection of primordial tensor modes~\cite{Akrami:2018odb},
it is natural to assume a low-energy small-field model represented by \eg hilltop inflation:
\bae{\label{eq: hilltop potential}
	V_{\mathrm{hill},i}(\phi_i)&=\left(v_i^2-g_i\frac{\phi_i^n}{\Mpl^{n-2}}\right)^2-\frac{1}{2}\kappa_i v_i^4\frac{\phi_i^2}{\Mpl^2}-\varepsilon_i v_i^4\frac{\phi_i}{\Mpl}, \nonumber \\ 
	n&\ge3,
}
for the phases responsible for PBHs. Here a dimensionful parameter $v_i$ determines the energy scale of phase-$i$ while $g_i$, $\kappa_i$, and $\varepsilon_i$ are dimensionless parameters controlling the duration of the phase, the shape of the scalar power spectrum, and so on.
$\Mpl=(8\pi G)^{-1/2}\simeq2.4\times10^{18}\,\mathrm{GeV}$ is the reduced Planck mass. We included the linear and quadratic terms as corrections to the simple wine-bottle potential.

In addition to this hilltop potential, we consider the following Planck-suppressed couplings between inflatons as stabilizers, 
the total potential being given by~\cite{Kawasaki:2016pql,
Inomata:2016rbd,Inomata:2017okj,Inomata:2017uaw,Inomata:2017vxo}
\bae{\label{eq: total potantial}
	V(\bm{\phi})=\sum_{i=1,2,3,4}V_{\mathrm{hill},i}(\phi_i)+\sum_{i\ne j}\frac{1}{2}c_{ij}V_{\mathrm{hill},i}(\phi_i)\frac{\phi_j^2}{\Mpl^2}.
}
Here we briefly review the dynamics of inflatons given by this potential. Let the energy scales be sufficiently hierarchical $v_1\gg v_2\gg v_3\gg v_4$.
Then the first inflationary phase is governed by $\phi_1$. During this phase, the high enough potential energy of $\phi_1$, $V_{\mathrm{hill},1}\simeq v_1^4$, stabilizes all other inflatons 
$\phi_2$, $\phi_3$, and $\phi_4$ to their origins through the couplings $\sum_{j\ne1}\frac{1}{2}c_{1j}V_{\mathrm{hill},1}\frac{\phi_j^2}{\Mpl^2}$.
After phase-1 inflation, $\phi_1$ rapidly oscillates around its potential minimum and loses its energy.
When the second energy scale $v_2^4$ becomes non-negligible compared to $\phi_1$'s energy, 
$\phi_2$ cannot be stabilized any longer and the second phase of inflation begins.
The same mechanism works for subsequent phases.
Without the linear term in the hilltop potential Eq.~(\ref{eq: hilltop potential}), the inflatons are stabilized completely at the origin and the dynamics after the onset of the phase
is stochastic, dominated by the quantum diffusion. Recalling that $V_{\mathrm{hill},i}\sim v_{i+1}^4$ at the beginning of phase-$(i+1)$,
the linear term shifts the minimum to $\phi_{i+1}\sim\phi_{*,i+1}=(\varepsilon_{i+1}/c_{i,i+1})\Mpl$ to make the background dynamics deterministic.\footnote{If the inflatons' dynamics is highly stochastic, one has to resort to \emph{stochastic-$\delta N$ formalism}~\cite{Fujita:2013cna,Fujita:2014tja,Vennin:2015hra} to calculate the power spectrum of the curvature perturbations.
However in this case, the curvature perturbations generally become too large and the resultant PBHs soon overclose the universe~\cite{Kawasaki:2015ppx}.}

After the stabilizer decays out, the potential tilt around this minimum reads $V_{\mathrm{hill},i+1}^\prime(\phi_{*,i+1})\sim-\varepsilon_{i+1}v_{i+1}^4/\Mpl$ 
and therefore 
the corresponding amplitude of the curvature power spectrum can be estimated as
\bae{
	\calP_{\calR}(k_{*,i+1})&\simeq\frac{1}{12\pi^2\Mpl^6}\frac{V_{\mathrm{hill},i+1}^3(\phi_{*,i+1})}{V_{\mathrm{hill},i+1}^{\prime2}(\phi_{*,i+1})} \nonumber \\
	&\sim\frac{1}{12\pi^2\Mpl^4}\frac{v_{i+1}^4}{\varepsilon_{i+1}^2}.
}
Hence $\varepsilon_{i+1}\sim v_{i+1}^2/\Mpl^2$ could realize enough power $\calP_\calR\sim\calO(10^{-2})$ for PBH formation as can be seen.
Apart from this peak, the $k$-dependence of the power spectrum is given by $\calP_\calR\propto k^{3-2\Re\nu}$ where the effective mass squared of $\phi_{i+1}$ determines
$\nu=\sqrt{\frac{9}{4}-\frac{m_{\eff,i+1}^2}{3H^2}}$. During phase-$i$, $\phi_{i+1}$ should be sufficiently massive as 
$m_{\eff,i+1}^2/3H^2\simeq c_{i,i+1}^2\gtrsim9/4$ to be well-stabilized, giving rise to the blue-tilted spectrum $\calP_\calR\propto k^3$ for $k<k_{*,i+1}$.
On the other hand, the $\phi_{i+1}$'s mass after the onset of phase-$(i+1)$ is given by $m_{\eff,i+1}^2/3H^2\simeq-\kappa_{i+1}$ 
and therefore large enough positive $\kappa_{i+1}$ can realize red-tilted spectrum for $k>k_{*,i+1}$. In this way, the power spectrum can have a sharp peak
on the onset scale $k_{*,i+1}$, and so does the corresponding PBH mass function~\cite{Kawasaki:1997ju,Kawasaki:1998vx}.\footnote{We note that in Refs.~\cite{Inomata:2016rbd,Inomata:2017vxo,Inomata:2017uaw} the sharp peak is realized mainly by the cancellation of the Hubble-induced mass during the oscillation phase due to the kinetic coupling, while it is simply a red-tilted one due to the inflaton's mass in this paper such as Refs.~\cite{Kawasaki:1997ju,Kawasaki:1998vx,Inomata:2017okj}. If one adopts the Press-Schechter approach with the Gaussian window which is relatively inefficient, the required primordial perturbations become large and the peak of the power spectrum should be highly sharp with use of such a complicated mechanism in order to avoid the current pulsar timing array (PTA) constraints on the stochastic GWs. However the real-space top-hat window used in this paper or the refined peak-theory approach~\cite{Yoo:2018esr} can reduce the accompanying GWs and be consistent with the current constraints without employing that mechanism.}
Large positive $\kappa_{i+1}$ also benefits us at the point of phase duration.
For the negligibly small first slow-roll parameter $\epsilon_H=-\dot{H}/H^2$ as we are considering, 
its time-evolution during the inflationary phase is controlled by the second derivative 
of the potential:
\bae{\label{eq: eHdot}
	\frac{\dot{\epsilon}_H}{H\epsilon_H}\simeq-2\Mpl^2\frac{V^{\prime\prime}_{\mathrm{hill},i+1}}{V_{\mathrm{hill},i+1}}\simeq2\kappa_{i+1}.
}
Therefore larger $\kappa_{i+1}$ shortens the duration of the inflationary phase-$(i+1)$, given the initial value of 
$\epsilon_H\simeq\frac{\Mpl^2}{2}\left(\frac{V^\prime_{\mathrm{hill},i+1}}{V_{\mathrm{hill},i+1}}\right)^2\simeq\frac{\varepsilon_{i+1}^2}{2}$
and the end of inflation $\epsilon_H=1$.
Multiple peaks in the PBH mass can be realized then.

\bfe{width=\hsize}{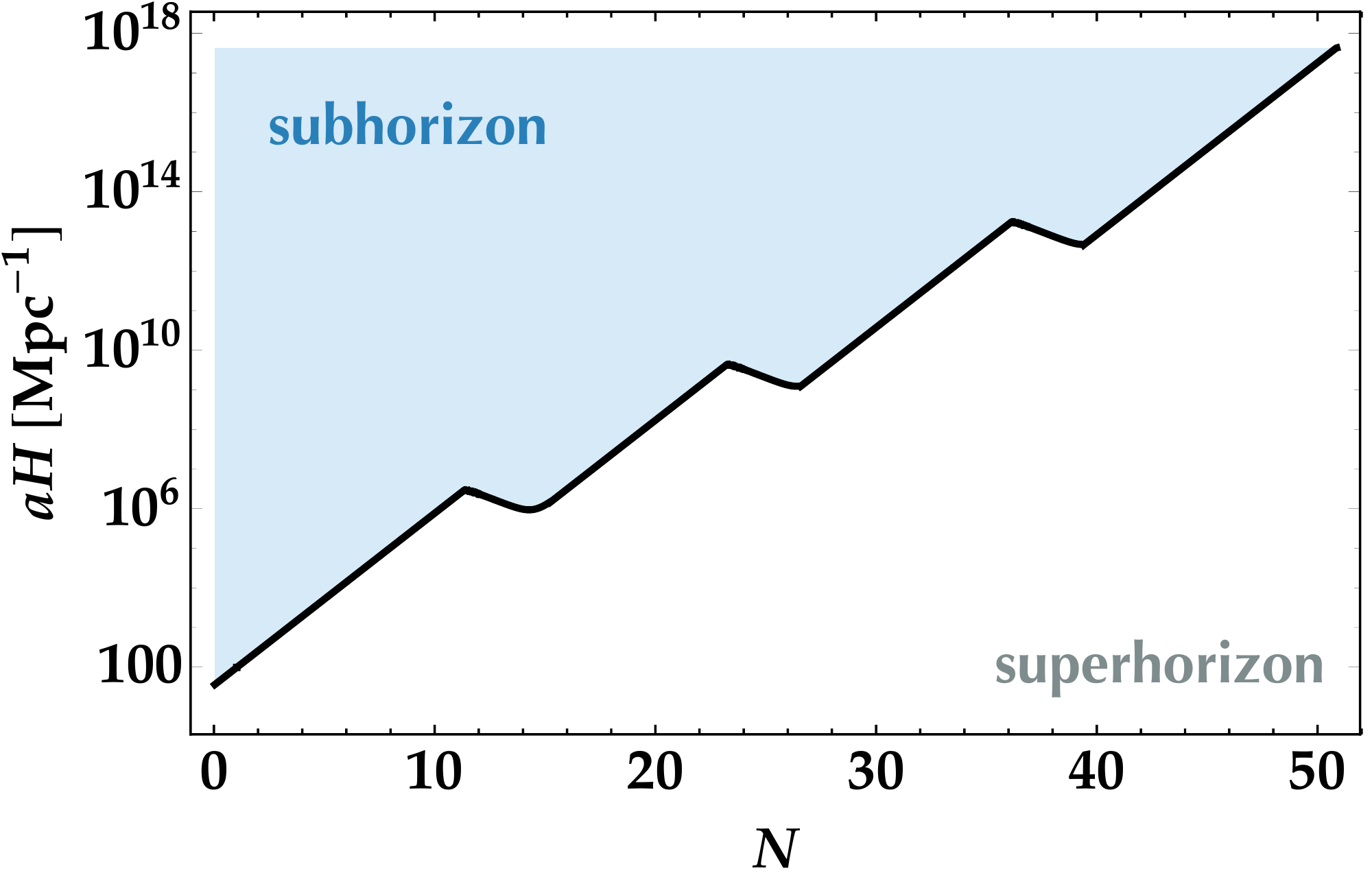}{The numerically obtained time (e-folds $N=\log a$) evolution of the horizon scale $aH$ under the parameters~(\ref{eq: params}).
The scale factor $a$ is consistently normalized by hand.
The quadruple inflationary ($aH$-growing) phases are shown, divided by the three oscillating ($aH$-decreasing) phases.}{fig: aH}

One finally takes account of the resonance condition at the end of hilltop inflation.
In general small-field inflation tends to bring about the resonance amplification of primordial perturbations soon after the end of inflation (see \eg Ref.~\cite{Brax:2010ai}),
which easily breaks the validity of the linear approximation of perturbations.
For a robust calculation, we avoid such a resonance in this paper.
The resonance condition of hilltop inflation has been investigated in detail in Ref.~\cite{Inomata:2017uaw},
and according to this work, roughly $n\le3$ and $\phi_{\mathrm{min},i}\simeq(v_i^2\Mpl^{n-2}/g_i)^{1/n}\gtrsim0.1\Mpl$ are favored.
Respecting this condition, we consider the following concrete set of parameters:
\bae{\label{eq: params}
	&n=3, \nonumber \\
	&\bce{
		\dps
		v_i/\Mpl&=(10^{-5},\,10^{-6},\,10^{-7},\,10^{-8}), \\ 
		\dps
		\varepsilon_i\Mpl^2/v_i^2&=(2,\,4.1,\,5.9,\,4.73), \\
		\dps
		\kappa_i&=(2,\,4.6,\,5,\,5), \\
		\dps
		(v_i^2\Mpl^{n-2}/g_i)^{1/n} &=0.1\Mpl, \qquad \text{for all $i$}, \\
		\dps
		c_{ij}&=10, \qquad \text{for all $i$ and $j$}.
	}
}
The numerically solved background dynamics is represented by the horizon scale $aH$ shown in Fig.~\ref{fig: aH}, 
successfully exhibiting quadruple phases of inflation. Then the Mukhanov-Sasaki equation~\cite{Mukhanov:1985rz,Sasaki:1986hm} gives us the full results of the linear perturbation as can be seen Fig.~\ref{fig: calPz}.
It consistently shows the three peaks corresponding with the horizon scales at the transition phases.
Moreover these peaks are well consistent with the dotted-line indicators $k^3$ and $k^{3-2\nu}$ where $\nu=\sqrt{9/4+\kappa_3}$ as discussed above.
It clarifies that there is no resonant amplification and the calculation of perturbations is well controlled.
For this power spectrum, the PBH mass spectrum shown in Fig.~\ref{fig: fPBH} is obtained.

\bfe{width=\hsize}{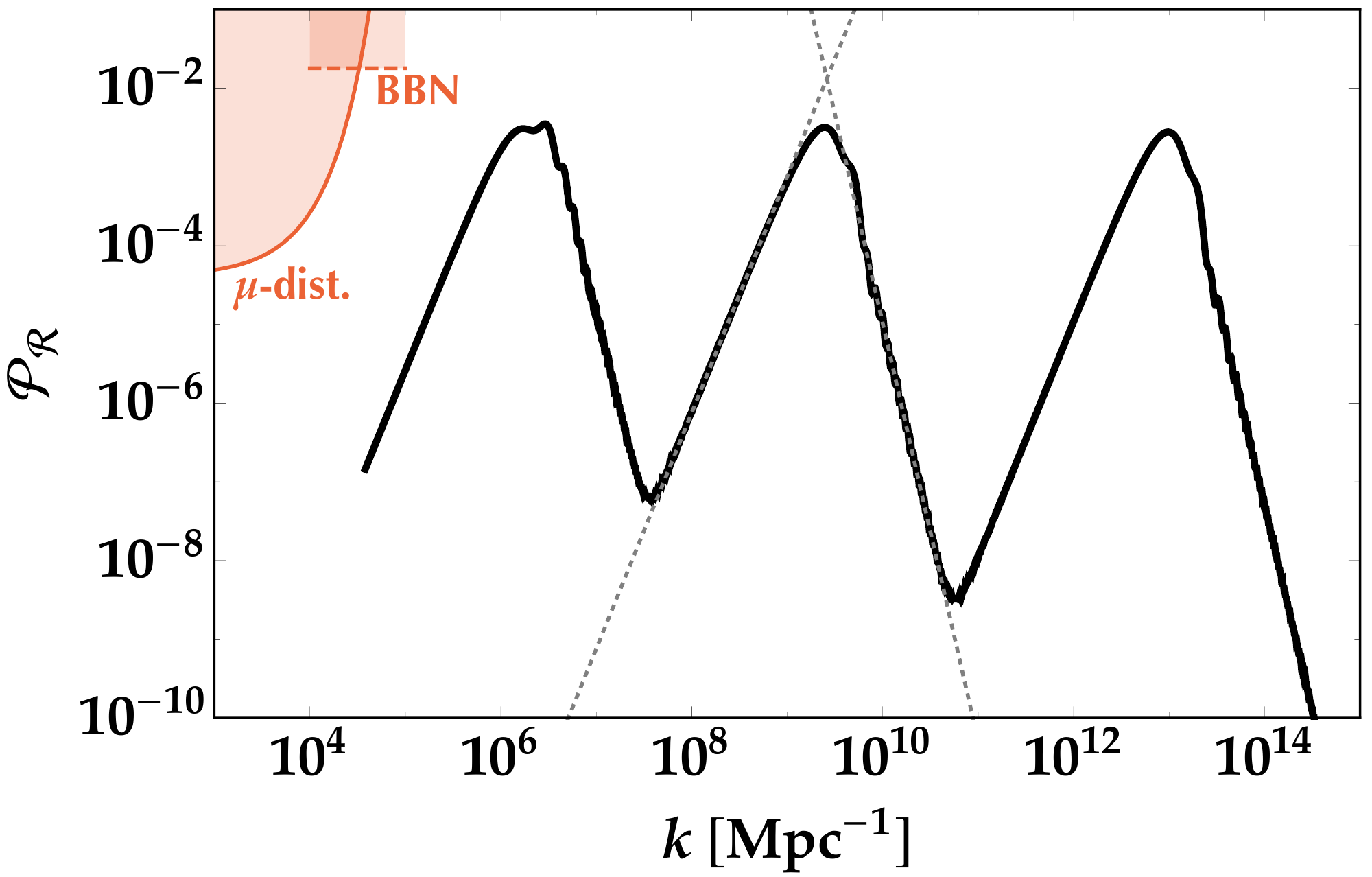}{The power spectrum of the curvature perturbations for the parameters~(\ref{eq: params}) obtained by solving the Mukhanov-Sasaki equation~\cite{Mukhanov:1985rz,Sasaki:1986hm}. Note that the normalization of the wavenumber $k$ is artificially fixed like as $aH$ in Fig.~\ref{fig: aH} by assuming a suitable cosmic evolution after phase-$4$. 
Consistent three peaks can be seen on the horizon scales ($k=aH$) at the transition phases shown in Fig.~\ref{fig: aH}.
Two dotted lines are indicators of $k^3$ and $k^{3-2\nu}$ where $\nu=\sqrt{(9/4+\kappa_3)}$ as the analytic prediction discussed in the main body.
Red-shaded regions show current upper bounds on the power spectrum: CMB $\mu$-distortion due to the Silk damping ($\mu$-dist.)~\cite{Chluba:2012we,Kohri:2014lza} and the effect on n-p ratio during big-bang nucleosynthesis (BBN)~\cite{Inomata:2016uip} (see also Refs.~\cite{Jeong:2014gna,Nakama:2014vla}).}{fig: calPz}

\section{Testability}\label{sec: GW}

As it has been discussed by many authors~\cite{Ananda:2006af,Baumann:2007zm,Saito:2008jc,Bugaev:2009zh,Saito:2009jt,Bugaev:2010bb,Byrnes:2018txb,Inomata:2018epa}, the inflationary PBH formation often leaves sizable stochastic GWs other than the binary PBH coalescence 
as a double-check evidence. Our model is also not an exception as we investigate in this section.

In the scenario of the collapsing radiation overdensity, abundant PBH formation requires the significant amplification of the primordial scalar perturbation as 
$\calP_\calR\sim10^{-2}$ under its Gaussian ansatz (see \eg Refs.~\cite{Young:2013oia,Nakama:2016kfq,Nakama:2016gzw,Cai:2018dig,Bartolo:2018rku,Unal:2018yaa} for the non-Gaussian effect on the required amplitude).
Even though the scalar and tensor perturbations are decoupled at the linear level, such a large amplitude of perturbations means the breakdown of
the linear approximation, so that the tensor mode (\ie stochastic GW) induced by the scalar perturbations becomes non-negligible.
Leaving the details below, one roughly obtains the following relation between the current density of induced GWs and the amplitude of 
the scalar perturbations:~\cite{Inomata:2016rbd}
\bae{\label{eq: Omega GW approx}
	\Omega_\mathrm{GW}h^2\sim10^{-9}\left(\frac{g_*}{10.75}\right)^{-1/3}\left(\frac{\Omega_rh^2}{4.2\times10^{-5}}\right)\left(\frac{\calP_\calR}{10^{-2}}\right)^2,
}
where $\Omega_rh^2\simeq4.2\times10^{-5}$ is the current radiative density parameter.
Because the current PTA constraints reach $\Omega_\mathrm{GW}h^2\sim10^{-9}$~\cite{Lentati:2015qwp,Shannon:2015ect,Aggarwal:2018mgp} and future GW-detection projects will go further,
it is required to predict the amplitude of the induced stochastic GWs in detail.

Taking the conformal Newtonian gauge:
\bae{
	\dd s^2&=-a^2(1+2\Phi)\dd\eta^2+a^2\left[(1-2\Psi)\delta_{ij}+h_{ij}\right]\dd x^i\dd x^j, \nonumber \\ 
	\partial_ih^i_j&=h^i_i=0,
}
the Einstein equation at the linear order in $h$ but the second order in $\Phi$ and $\Psi$ can be summarized by~\cite{Inomata:2016rbd}
\bae{
	h_{ij}^{\prime\prime}+2\calH h_{ij}^\prime-\nabla^2h_{ij}=-4\hat{\calT}_{ij;kl}S_{kl}, \qquad \calH=\frac{a^\prime}{a},
}
with the source term
\bae{
	S_{ij}=&\,4\Psi\partial_i\partial_j\Psi+2\partial_i\Psi\partial_j\Psi \nonumber \\
	&-\frac{4}{3(1+w)}\partial_i\left(\frac{\Psi^\prime}{\calH}+\Psi\right)
	\partial_j\left(\frac{\Psi^\prime}{\calH}+\Psi\right).
}
Here $\hat{\calT}_{ij;kl}$ represents the traceless-transverse projection operator satisfying
\bae{
	\hat{\calT}_{ij;kl}\hat{\calT}_{kl;mn}=\hat{\calT}_{ij;mn}, \qquad \partial_i\hat{\calT}_{ij;kl}\calO_{kl}=\hat{\calT}_{ii;kl}\calO_{kl}=0,
}
for any second-order tensor $\calO_{kl}$. $w=p/\rho$ is the background equation of state, and we assume the negligible anisotropic stress as $\Phi=\Psi$.

In terms of the conserved curvature perturbation $\calR$, the time average of the power spectrum of induced GW for each polarization mode is expressed by
\bae{\label{eq: calPh}
	&\hspace{-10pt}\overline{\calP_h(\eta,k)} \nonumber \\
	=&\,2\int^\infty_0\dd t\int^1_{-1}\dd s\left(\frac{t(2+t)(s^2-1)}{(1-s+t)(1+s+t)}\right)^2\overline{I^2(s,t,k\eta)} \nonumber \\
	&\times\calP_\calR\left(k\frac{1-s+t}{2}\right)\calP_\calR\left(k\frac{1+s+t}{2}\right),
}
and the Kernel function $I^2$ can be approximated in the subhorizon limit $x=k\eta\to\infty$ during the radiation-dominated universe 
as~\cite{Inomata:2018epa,Kohri:2018awv}
\begin{widetext}
\bae{
	\overline{I^2(s,t,x\to\infty)}\simeq&\,\frac{288(-5+s^2+t(2+t))^2}{x^2(1-s+t)^6(1+s+t^6)}\left[\frac{\pi^2}{4}(-5+s^2+t(2+t))^2\Theta(t-\sqrt{3}+1) \right. \nonumber \\
	&\left.+\left(-(1-s+t)(1+s+t)+\frac{1}{2}(-5+s^2+t(2+t))\log\left|\frac{-2+t(2+t)}{3-s^2}\right|\right)^2\right],
}
\end{widetext}
where $\Theta$ is the Heaviside step function.
The energy density of GWs per logarithmic $k$-bin is related to the power spectrum by
\bae{
	\Omega_\mathrm{GW}(\eta,k)=\frac{1}{24}\left(\frac{k}{aH}\right)^2\overline{\calP_h(\eta,k)}.
}
Making use of the relation $aH=\eta^{-1}$ during the radiation-dominated era, one finds that the time dependence in $\Omega_\mathrm{GW}$ is cancelled 
soon after its horizon reentry. After this freezeout time $\eta_c$, the density ratio of GWs to the background radiation is almost constant up to the current time $\eta_0$.
Therefore, taking account of the change of the effective degrees of freedom, the current density parameter for the induced GWs is given by~\cite{Inomata:2018epa}
\bae{\label{eq: OmegaGW now}
	\Omega_\mathrm{GW}(\eta_0,k)=0.83\left(\frac{g_*}{10.75}\right)^{-1/3}\Omega_r\Omega_\mathrm{GW}(\eta_c,k).
}
For a monochromatic scalar power spectrum $\calP_\calR=A\delta(\log k-\log k_*)$, the GW energy spectrum has a peak on $k_p=2k_*/\sqrt{3}$ and 
the integration over one $\log k$ bin around this peak scale $\log k_p-1/2<\log k<\log k_p+1/2$ reproduces the approximated estimation 
formula~(\ref{eq: Omega GW approx}).

In Fig.~\ref{fig: OmegaGW}, we show the numerical result of predicted energy density of secondary GWs induced by our scalar power spectrum shown in 
Fig.~\ref{fig: calPz}, with use of Eqs.~(\ref{eq: calPh}--\ref{eq: OmegaGW now}).
The current observational constraints and the expected sensitivities of future projects are also shown by several lines with/without shades respectively.
While the predicted GWs evade all current constraints,
they can be tested \eg by SKA for the stellar-mass PBHs ($k\sim10^6\,\mathrm{Mpc}^{-1}$) and by several space-based detectors for DM PBHs with the asteroid-mass ($k\sim10^{12}\,\mathrm{Mpc}^{-1}$).
The earth-mass PBHs ($k\sim10^9\,\mathrm{Mpc}^{-1}$) might be marginally detectable with LISA, depending on the estimation procedure of the PBH abundance.

\bfe{width=\hsize}{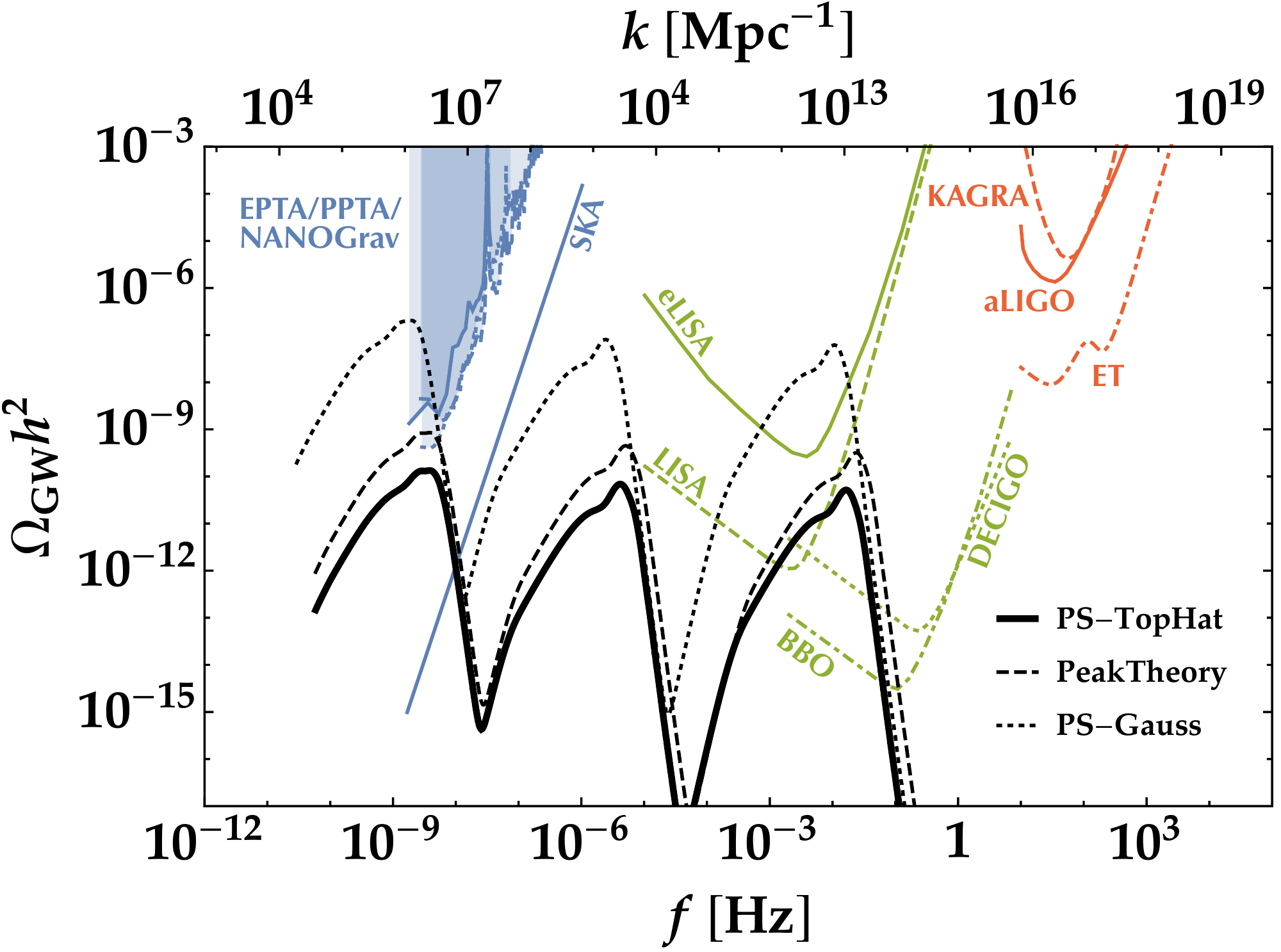}{The predicted current abundance of the stochastic GWs in our model (Black thick line) and the sensitivities of the current/future projects for the GW detection.
Blue lines represent the pulsar timing array sensitivities: the current constraints by EPTA~\cite{Lentati:2015qwp}, PPTA~\cite{Shannon:2015ect}, and NANOGrav~\cite{Aggarwal:2018mgp}, and the prospective sensitivity of SKA~\cite{Moore:2014lga}.
The sensitivities of the space-based projects (eLISA, LISA, DECIGO, and BBO) are summarized in Ref.~\cite{Moore:2014lga}. Red lines indicate the ground-based ones: aLIGO and KAGRA~\cite{Aasi:2013wya}, and the Einstein Telescope (ET)~\cite{Sathyaprakash:2009xs}.
Black dashed/dotted lines correspond with other estimation procedures of the PBH abundance. See Appendix~\ref{sec: uncertainties} for details.}{fig: OmegaGW}

\section{Discussion and Conclusions}\label{sec: conclusions}

In this paper, we discussed the PBH formation with the hierarchical mass function in a multi-phase inflationary model.
Once one allows the multiple phase for the last $50\text{--}60$ e-folds, the power spectrum of the primordial curvature perturbation can easily have peaks on the transition scales,
as we showed by considering a simple model: four single-field hilltop inflations~(\ref{eq: hilltop potential}) coupled by the natural Planck-suppressed terms~(\ref{eq: total potantial}).
With the parameters~(\ref{eq: params}) chosen to avoid the resonant amplification for simplicity, the linear perturbation theory consistently gives the peaky power spectrum shown in Fig.~\ref{fig: calPz}. Following the Press-Schechter approach, one obtains the current expected abundance of PBHs as Fig.~\ref{fig: fPBH}. Here this mass spectrum satisfies three interesting motivations of PBHs simultaneously, that is, $\calO(10)M_\odot$ BHs for the LIGO/Virgo GW events, $\calO(10^{-12})M_\odot$ ones as a main component of DMs, and as recently pointed out~\cite{Niikura:2019kqi}, $\calO(10^{-5})M_\odot$ PBHs explaining the OGLE ultrashort-timescale microlensing events.
Large primordial scalar perturbations on the other hand yield sizable stochastic GWs through the second order effect. Fig.~\ref{fig: OmegaGW} shows the corresponding GW abundance with the power spectrum in Fig.~\ref{fig: calPz}, compared with several current/future sensitivities of GW detectors.
As discussed in Appendix~\ref{sec: uncertainties}, the uncertainties in the theoretical estimation of the PBH abundance cause non-negligible differences in the prediction of the GW abundance. This issue is important for future prospects.

We also mention the recent de Sitter swampland conjecture in the context of string theory.
Reflecting the difficult situations so far to realize de Sitter vacua in the string theory, 
Ooguri, Vafa, and other authors have rather conjectured that any de Sitter like state will be unstable in consistent theories of quantum gravity~\cite{Obied:2018sgi,Garg:2018reu,Ooguri:2018wrx}.
The concrete statement is as follows.
\begin{swampland}
For any consistent quantum gravity,
a scalar potential $V(\phi)$ for its low energy effective field theory denoted by the effective action
\bae{
	S\!=\!\!\int\!\dd^4x\sqrt{-g}\left[\frac{1}{2}M_\mathrm{Pl}^2R\!-\!\frac{1}{2}g^{\mu\nu}G_{I\!J}(\phi)\partial_\mu\phi^I\partial_\nu\phi^J\!-\!V(\phi)\right]\!,
}
must satisfy either
\bae{
	|\nabla V|M_\mathrm{Pl}\ge cV,
}
or
\bae{
	\min(\nabla_I\nabla_JV)M_\mathrm{Pl}^2\le-c^\prime V,
}
at any field-space point for some universal constants $c,\,c^\prime>0$ of order unity.
Here $\nabla_I$ is the covariant derivative along with the field-space metric $G_{IJ}$, $|\nabla V|=\sqrt{G^{IJ}\nabla_IV\nabla_JV}$ is 
the invariant norm of the potential tilt, and $\min(\nabla_I\nabla_JV)$ is the minimum eigenvalue of the Hessian $\nabla_I\nabla_JV$ in an orthonormal frame.
\end{swampland}
In terms of inflation, this conjecture claims that any single continuous inflationary phase cannot last so long.
That is, if it is true, one needs multiple phases of inflation to explain sufficient e-folds $\sim50\text{--}60$ for our observable universe in total,
as we proposed.
In fact our model always satisfies the second condition during any inflation phase as
\bae{
	-\frac{\min(\nabla_I\nabla_JV)}{V}\simeq\kappa_i>1, \qquad \text{for all phase-$i$},
}
while the first condition is satisfied apart from the inflationary trajectory.
The existence of large negative eigenvalue in the Hessian matrix generally causes a negligibly short (less than 1 e-fold) inflationary phase
unless the initial value of the first slow-roll parameter $\epsilon_H$ is significantly small (see Eq.~(\ref{eq: eHdot})), 
so that the corresponding amplitude of the power spectrum is large.
In other words, the PBH formation on the onset scale is relatively natural for a non-negligibly continuing phase in the context of the above conjecture.

\acknowledgments

We are grateful to Keisuke Inomata and Chul-Moon Yoo for useful discussions.
YT is supported by Grant-in-Aid for JSPS Research Fellow (JP18J01992).
SY is supported by MEXT KAKENHI Grant Number 15H05888 and
18H04356.

\appendix

\section{Uncertainties in PBH abundance prediction}\label{sec: uncertainties}

Even though the statistics of primordial curvature perturbations are completely fixed,
the analysis of PBH formation of the fully non-linear process requires some approximations.
That causes theoretical uncertainties in the prediction of the PBH abundance.
For example, the Press-Schechter approach characterizes overdensities simply by the universal threshold value and the typical window function. However of course profiles of overdensities are not uniform, varying the corresponding threshold value as $\delta_\thr\sim0.4\text{--}0.6$~\cite{Musco:2004ak}.
Moreover, while it naively assumes the one-to-one correspondence between the coarse-graining scale $R$ and the PBH mass $M(R)$ with a single parameter
$\gamma=M(R)/\left.\left(\rho\frac{4\pi}{3}H^{-3}\right)\right|_{aH=R^{-1}}$ (\ref{eq: MR}),
it is known that the resultant PBH mass depends on the shape of the density profile and the excess density (see \eg Refs.~\cite{Choptuik:1992jv,Niemeyer:1999ak,Musco:2004ak,Musco:2008hv,Musco:2012au,Kuhnel:2015vtw}).
Given the PBH mass spectrum inversely, such uncertainties vary the required primordial power spectrum of curvature perturbations and thus the prediction of the current energy density of the secondary gravitational waves (GWs)~\cite{Ando:2018qdb}.
In Fig.~\ref{fig: OmegaGW}, we show the estimated GW density by the same PBH mass spectrum in Fig.~\ref{fig: fPBH} but with the different choice of parameters as $\delta_\thr=0.6$, $\gamma=w=1/3$, and the Gaussian window function:
\bae{
    \tilde{W}_G(kR)=\exp\left(-\frac{(kR)^2}{2}\right),
}
by the black dotted line.\footnote{As remarked in Ref.~\cite{Ando:2018qdb}, the coarse-grained volume slightly depends on the choice of the window function as $V(R)=4\pi R^3/3$ for the real-space top-hat while $V(R)=(2\pi)^{3/2}R^3$ for the Guassian window. This gives rise to the difference in the corresponding PBH mass.}
Compared to the top-hat one, subhorizon modes do not contribute to the variance $\sigma_\delta^2(R)$~(\ref{eq: sigma_delta}) with the Gaussian window, so that larger primordial curvature perturbations are required even for the same abundance of PBHs, together with the high threshold choice $\delta_\thr=0.6$.
Accordingly, in this selection of parameters, our PBH mass spectrum is in tension with the current PTA constraints on GWs.

Recently Yoo \etal proposed a novel estimation procedure of the PBH abundance in the peak theory~\cite{Yoo:2018esr} (see also Ref.~\cite{Germani:2018jgr}
where peaks in density perturbations are used instead of curvature perturbations).
The formation of extremely rare objects such as PBHs can be described in the high peak limit.
According to the peak theory~\cite{Bardeen:1985tr}, the radial profile of such a high peak of Gaussian curvature perturbations can be characterized by the typical form
parametrized by stochastic variables.
Therefore the PBH formation can be discussed statistically without ambiguity.
Particularly the PBH mass is stochastically determined by the curvature of the profile in principle and therefore any specific window function needs not to be introduced.
However this formulation requires the power spectrum of curvature perturbations to have an exponentially sharp single peak 
because otherwise the profile is contaminated by other wavelength modes.
Since the tails of our peaks decay only in the power law, our model is slightly out of the application scope of this approach, and therefore we simply apply the approximated estimation of the peak value of the mass spectrum (65) of Ref.~\cite{Yoo:2018esr} to the three peaks of our curvature power spectrum.
Let us briefly review this estimation below.

Thanks to the high peak limit, the PBH formation is mainly contributed by the peak scale mode $k_c$ determined by
\bae{
	k_c=\frac{\sigma_1}{\sigma_0}, \qquad \sigma_n^2=\int\frac{\dd k}{k}k^{2n}\calP_\calR(k).
}
For this mode, the profile of the peak curvature perturbation is simply given by the two-point function
\bae{
    \psi(r)=\frac{1}{\sigma_0^2}\int\frac{\dd k}{k}\frac{\sin kr}{kr}\calP_\calR(k),
}
as\footnote{Our definition of the curvature perturbation has an opposite sign from Ref.~\cite{Yoo:2018esr} so that the positive $\calR$ coincides with the positive excess in the spatial curvature.}
\bae{\label{eq: curvature profile}
    \calR(r)=\mu\psi(r), \qquad \mu=\calR(r=0).
}
If the maximal value of the compaction for this profile exceeds some threshold, the corresponding peak is assumed to collapse.

The compaction function is defined by
\bae{
    \calC(r)=\frac{\delta M}{R},
}
where $\delta M$ represents the excess of the Misner-Sharp mass from the expected one by the background universe and $R$ is the areal radius. It can be simplified as
\bae{
    \calC(r)=\frac{1}{3}\left[1-(1+r\calR^\prime)^2\right],
}
and therefore the maximally compact radius $r_m$ can be found by the condition
\bae{
    \calC^\prime(r_m)=0, \quad \Leftrightarrow \quad \calR^\prime+r\calR^{\prime\prime}|_{r=r_m}=0.
}
The threshold value for the compaction is estimated as $\calC_\thr\simeq0.267$~\cite{Yoo:2018esr}. Therefore the PBH formation criterion is given by
\bae{
    \calC(r_m)\ge\calC_\thr, \quad \Leftrightarrow \quad \mu\ge\mu_c=\frac{\sqrt{1-3\calC_\thr}-1}{r_m\psi^\prime(r_m)}.
}
With use of this threshold, 
the estimated PBH density ratio $\beta$ to the background at its formation time is given by the Gaussian distribution with the phase-space volume factor $\mu_c^3/\sigma_0^3$ as\footnote{Note that the definition of $\beta$ is different from that in the main body~(\ref{eq: betaR}) by the factor $\gamma$.}
\bae{
    \beta\sim\gamma\frac{\mu_c^3}{\sigma_0^3}\ee^{3\mu_c\psi(r_m)}
    \exp\left[-\frac{\mu_c^2}{2\sigma_0^2}\right].
}
Here the factor $\ee^{3\mu_c\psi(r_m)}$ comes from the ratio of the physical volume with the areal radius $R$ and the background value corresponding with the coordinate radius $r_m$. The PBH mass is given by the horizon mass~(\ref{eq: MR}) as
\bae{
    M&=\left.\gamma\rho\frac{4\pi}{3}H^{-3}\right|_{H=R^{-1}} \nonumber \\
    &\simeq10^{20}\gamma\left(\frac{g_*}{106.75}\right)^{-1/6}\left(\frac{r_m\ee^{\mu_c\psi(r_m)}}{6.4\times10^{-14}\,\mathrm{Mpc}}\right)^2\,\mathrm{g},
}
and the current fraction to total DMs~(\ref{eq: fPBH}) reads
\bae{
    &f_\mathrm{PBH} \nonumber \\
    &\simeq\!\gamma^\frac{1}{2}\!\left(\!\frac{\beta}{7.2\times10^{-16}}\!\right)\!\left(\!\frac{\Omega_\mathrm{DM}h^2}{0.12}\!\right)^{\!-1}\!\biggl(\!\frac{g_*}{106.75}\!\biggr)^{\!-\frac{1}{4}}\!\left(\!\frac{M}{10^{20}\,\mathrm{g}}\!\right)^{\!-\frac{1}{2}}\hspace{-5pt}.
}

We then plot the corresponding secondary GW spectrum with $\gamma=1$ by the black dashed line in Fig.~\ref{fig: OmegaGW}, fixing the three peaks of the PBH mass function.
Because this improved estimation procedure is more efficient, the resultant GW density becomes lower than that for the Press-Schechter approach with the Gaussian window.
Consequently it is marginally consistent with the current PTA constraints.

\bibliography{main}
\end{document}